\documentclass[aps,prl,twocolumn]{revtex4}
\usepackage{amssymb,amsmath,graphicx,color,microtype}
\usepackage{graphicx}
\usepackage{txfonts}

\usepackage{epsf}
\usepackage{epstopdf}
\usepackage{amssymb}
\newcommand{\nc}{\newcommand}
\newcommand\be{\begin{equation}}
\newcommand\ee{\end{equation}}

\usepackage{float}

\nc{\e}{{\bf{e}}}
\nc{\kk}{{\bf{k}}}
\nc{\pp}{{\bf{p}}}

\nc{\bfk}{{\bf{k}}}
\nc{\bfx}{{\bf{x}}}
\nc{\bfp}{{\bf{p}}}

\nc{\eH}{{\epsilon_H}}
\nc{\calP}{{\cal P}}
\nc{\im}{{ \mathrm{Im} } }

\def\apj{Astrophysical Journal}
\def\mnras{Monthly Notes of Royal Astronomical Society}

\def\araa{Annual Review of Astronomy \& Astrophysics}

\def\aap{Astronomy and Astrophysics}     
\def\prd{Phyical Review D}
\def\prl{Phyics Review Letter}

\begin{document}

\author{Tom~Broadhurst}
\affiliation{Department of Theoretical Physics, University of the Basque Country UPV-EHU, 48040 Bilbao, Spain}
\affiliation{Donostia International Physics Center (DIPC), 20018 Donostia, The Basque Country}
\affiliation{IKERBASQUE, Basque Foundation for Science, Alameda Urquijo, 36-5 48008 Bilbao, Spain}

\author{Jose~M. Diego}
\affiliation{Instituto de F\'isica de Cantabria, CSIC-Universidad de Cantabria, E-39005 Santander, Spain}

\author{George~F.~Smoot}

\affiliation{Donostia International Physics Center (DIPC), 20018 Donostia, The Basque Country}
\affiliation{ IAS TT \& WF Chao Foundation Professor, IAS, Hong Kong University of Science and Technology,
Clear Water Bay, Kowloon, 999077 Hong Kong, China}
\affiliation{Paris Centre for Cosmological Physics, Universit\'{e} de Paris, CNRS,  Astroparticule et Cosmologie, F-75013 Paris, France A, 10 rue Alice Domon et Leonie Duquet,
75205 Paris CEDEX 13, France, {\it Emeritus}}
\affiliation{Physics Department and Lawrence Berkeley National Laboratory, University of California, Berkeley,
94720 CA, USA, {\it Emeritus} }

\title{A uniform stellar origin for binary black holes  revealed by lensing}






\begin{abstract}

 Although most gravitational wave events are claimed to be mergers of unusually massive, $25-65M_\odot$, black holes, it is now clear that 20\% of all reported events comprise modest mass black holes, $5-15M_\odot$, like the stellar black holes in the Milky Way. We show that such stellar mass black hole binaries (BBH) if magnified by lensing galaxies can be detected at high redshift, 1$< $z$ <$5, with chirp masses increased by $1+z$, accounting for the majority of apparently high mass BBH events.
 This simple lensing explanation is manifested by the evident bimodality of BBH chirp masses now visible, with 80\% of BBH events in a broad peak centered on $m_{chirp} \simeq 35M_\odot$, and 20\% of BBH events 
 in a narrow, low mass peak at $m_{chirp} \simeq 8.5M_\odot$, matching well our prediction for lensed and unlensed events respectively. This lensing interpretation is reinforced by the ``graveyard plot" when ranked by chirp mass, revealing a jump in chirp mass at $m_{chirp} \simeq 10M_\odot$ that we show is caused by the large redshift difference between unlensed events with $z<0.3$ and lensed events above $z>1$. Furthermore, nearly all BBH events are seen to share a component mass ratio of $m_1/m_2=1.45\pm0.03$, indicating a common stellar origin for BBH events across all chirp masses. This observed component mass uniformity implies most binary black holes seldom pair up by random capture, instead we may conclude that massive progenitor stars of BBH black holes typically formed in-situ, in a well defined way over the full span of cosmic time accessed through gravitational lensing.
  
\end{abstract}




\maketitle



   Decades of EM-observations have established that black holes orbiting nearby stars have a narrow range of masses, $5-15M_\odot$, peaking at $8M_\odot$, with none known above $20M_\odot$, or below about 5$M_\odot$\cite{Remillard,El-Badry,Stanway}. Most local stellar black holes were identified by X-ray gas emission but more recently new methods have uncovered a black hole of $11.1\pm 2.2 M_\odot$ from direct orbital motion \cite{Saracino2021}, and a micro-lensing black hole of $7.1\pm1.3 M_\odot$ \cite{Sahu2022} has been determined from stellar astrometry, thereby independently reinforcing the stellar black hole mass mass range, $5-15M_\odot$. Here we show that this stellar black hole mass range is  shared by all 30 black holes that comprise the 15 lowest chirp mass BBH detections reported to date, which also span the range, $5-15M_\odot$, implying that these low chirp mass events comprise pairs of conventional black holes where both members are remnants of massive stars. This stellar mass population is clear in Figure~1 where we have ranked all BBH events by chirp mass, showing these 15 lowest mass events are centered on a chirp mass of $\simeq 8.5M_\odot$ and appear distinct form the majority of events of higher chirp mass. This distinction is also apparent in the component mass distribution of observed black hole masses shown in Figure~3, where a sharp, low mass peak is distinct from the majority of black holes that span a broad mass distribution centered on $\simeq 35M_\odot$, which we discuss below in the context of lensing. 
  
\begin{figure*}[ht]
	\centering
\includegraphics[width=17.5cm,height=10cm]{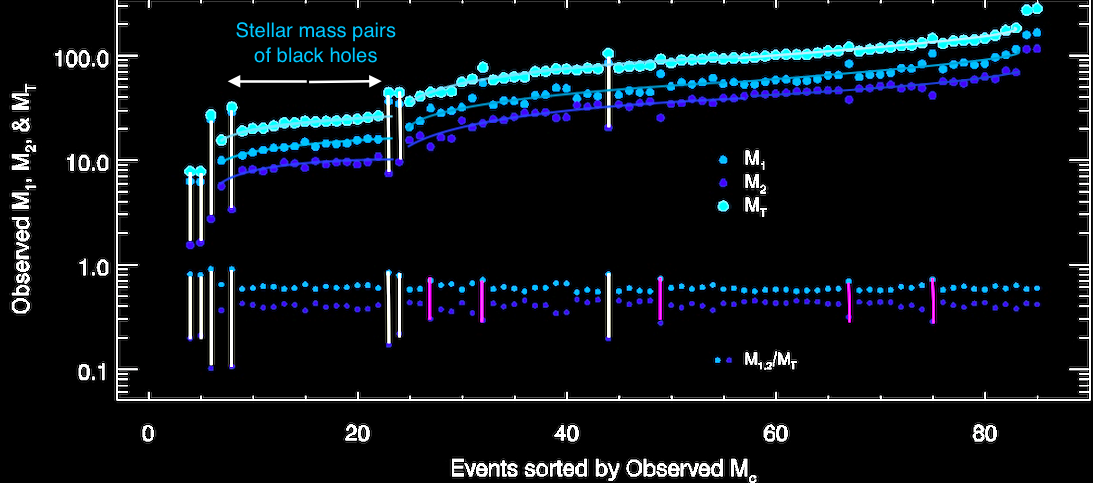} 

  
  
\caption{{\bf All reported binaries ranked by chirp mass} This ranked version of the ``Graveyard" plot reveals that BBH events share a common component mass ratio, forming a pair of remarkably parallel tracks when normalised by the total binary mass (lower point set), defining a common mass ratio of $1.45\pm0.03$ that spans the full range of chirp mass. We can also see here a step-like transition between the ``stellar mass" group of 15 lowest chirp mass BBH events centered on a rest frame value of $8.5M_\odot$, with $z<0.3$, and all the higher chirp mass events that follow a slowly rising curve, compatible with being high redshift, lensed examples of the lower mass ``stellar mass" group. The commonality in mass ratio between high and low chirp mass BBH events implies a common origin. Events marked as white lines with large mass ratios are the reported NSBH events and the lower Mass-Gap and Mass-Asymmetric events. The events marked in purple are discussed as possible second generation BBH mergers.
}
\label{r0-value2}
\end{figure*}

\begin{figure}[ht]
	\centering
\includegraphics[width=1.0\columnwidth,height=13cm]{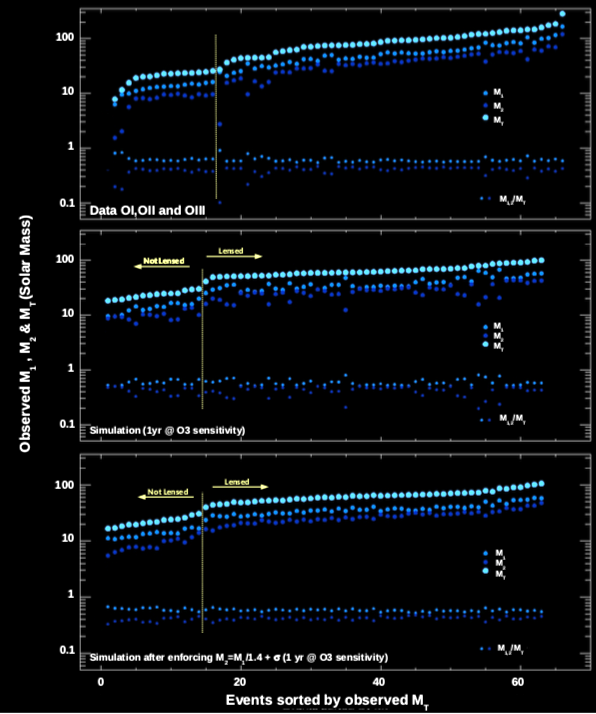} 
  
  
\caption{{\bf Comparison with Lens Model}. Predictions from our lens model for one year of observation at O3 sensitivity in the central and low panels, for comparison with the most trusted events reported by the LVK team in the upper panel. The predictions show a clear distinction between unlensed and lensed events marked with a vertical dashed line, arising directly from the high redshift enhancement of the chirp masses by $1+z$ for lensed events compared to the predicted unlensed events, for which $z<0.3$ - like the reported redshifts of the low chirp mass events. In the central panel we allow the component masses to be drawn randomly from the input lognormal mass distribution which has a wider spread of component mass ratios than in the upper panel of the data,  whereas, in the lower panel matches well the mass ratios of the data when we restrict the component mass ratio to $m_1/m_2=1.45$ with a small dispersion of $\simeq$ 10\%.}
\label{r0-value2}
\end{figure}
 
   It is also striking in Figure~1 that nearly all BBH events of high and low mass appear to share a common characteristic mass ratio of $m_1/m_2\simeq 1.45$, resulting in two very uniform, parallel bands when normalising by the total binary mass that span the full chirp mass range of Figure~1. To interpret this uniformity in mass ratio, it should first be appreciated that the cosmological expansion of GW waveform by $1+z$ exactly mimics a higher chirp mass binary \cite{Wang,Broadhurst2018} since gravitation is scale free, so binaries with masses $(m_1, m_2)$ in the rest frame are indistinguishable by frequency dependence from masses:
    $(\frac{m_1}{1 + z} , \frac{m_2}{ 1 + z})$, at redshift $z$, as the cosmological time dilation ensures the waveform is shifted by precisely the necessary amount to preserve the chirp shape and frequency by the same factor of $1+z$ for both the primary and secondary masses. Hence, the uniformity of the component mass ratios is inherent to the BBH black holes and independent of their redshifts. 
   
   We now draw attention to the striking transition visible in Figure~1, marked by an upward step in chirp mass from $10.5M_\odot$ to $15.5M_\odot$, above the range of stellar mass black holes, indicated in figure~1 between the 15 low mass events which we have argued are stellar mass BBH binaries at low redshift, and all higher chirp mass events. In the context of lensing this transition neatly distinguishes high-z lensed events from low-z unlensed events because the chirp mass scales with redshift as $1+z$ via:
 \begin{equation}
     M_{\rm Chirp}(z)=(1+z)M_{\rm Chirp}(0)
\end{equation}
   Hence, this observed transition in chirp mass 
   by $\simeq 5M_\odot$ is equivalent to an increase in $1+z$ of $\simeq 50\% $ which translates into a minimum redshift of $z>1.4$ for events above this transition relative to the measured mean redshift of $\bar{z}\simeq 0.2$ reported for the stellar mass BBH events lying below this transition. 
  
\begin{figure}[ht]
 \centering
 \vspace{-1pt}

  \includegraphics[width=1.0\columnwidth,height=8cm]{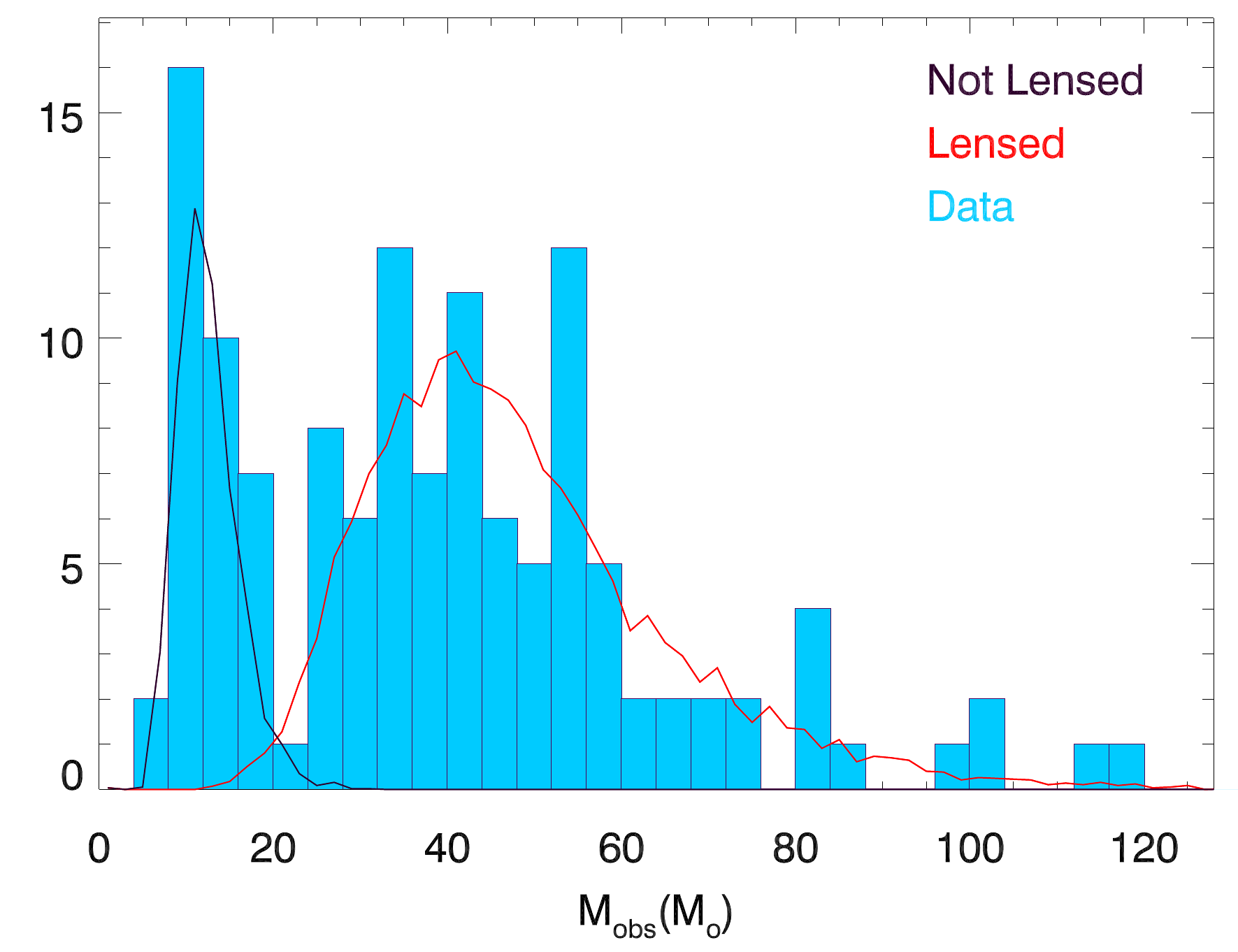} 
  \hspace{-20pt}
  \vspace{-1pt}
\caption{{\bf Distribution of binary black hole masses} Histogram of all 134 individual black hole masses for the 67 BBH events reported to date above
an SNR limit of 8 that we have modelled. The red curve shows our model prediction for lensed events and the black curve is for unlensed events. Both predicted curves have characteristic peaks because they derive from the same, relatively narrow log-normal distribution of the known stellar black holes in our Galaxy and together describe a bimodal distribution that matches the data well, with a mean redshift of $z\simeq 0.2$ and $z\simeq 2.8$ for unlensed and lensed BBH events respectively.}
\label{r0-value2}
\end{figure}

Regarding the origin of the high chirp mass events, we have argued that above a chirp mass of $\simeq 15M_\odot$, all events are gravitationally lensed, because BBH binaries with normal stellar mass black holes of $5-15M_\odot$ are detectable in the range $1<z<5$ at the current instrumental limits if lensed, with chirp masses dominated by the $1+z$ expansion factor (eqn.~1). We now apply our established lens model to generate distributions of lensed BBH events, based on the simplest, empirically based a-priori assumptions, by using the known mass function for stellar black holes and adopting the universal form for the high magnification tail of fold caustics, $\propto \mu^{-3}$ normalised by the known optical depth of galaxy lensing from sky surveys. Lensing by galaxies or clusters does not of course alter the frequencies of gravitational waves, but the detected strain is magnified, such that the inferred distance will be substantially underestimated by $\sqrt{\mu}$: 

\begin{equation}
D_L(z_{inferred}) =\frac{{\rm M}_{chirp, z}^{5/6}}{h(t)} ~ F(t,{\rm M}_{obs},\Theta) = \frac{D_L(z_{true})}{\sqrt{\mu}}
\label{Eq_ht}
\end{equation}

 For binary events the detection function $F(t,{\rm M}_{obs},\Theta)$ combines the angular sky sensitivity, orbital inclination, spin and polarization of the binary source\cite{Finn} and its distribution is numerically estimated with a $\simeq 40\%$ dispersion\cite{Ng,Abbot_2016}. Our only major unknown is the evolution of BBH events for which there is a wide range of ideas\cite{Sig,Ban,Ugo,Farr}, so we adopt an exponentially declining BBH event rate for $z<2$ as the simplest assumption that we find to reproduce  the approximate 4-to-1 ratio of lensed to unlensed BBH events at the current depth of the data, as shown in Figures 1 \& 2, corresponding to an e-folding time of $\simeq 0.8 Gyr$ (see appendix). We stress that the rate of evolution is effectively the only degree of freedom we have in our lens model because the input mass function and the lensing optical depth are set by real priors established before GW events were detected, as outlined in detail in our first calculations\cite{Broadhurst2018} which we refer to as the BDS model. 

Firstly, we show our predictions for unlensed BBH events with the BDS model, which depends mainly on the input BBH component mass distribution and not on lensing. Our predicted unlensed events span the range $z<0.3$ for the current O3 sensitivity, with a peak in detections at a chirp mass of $8-10M_\odot$. This is very similar to the inferred redshifts and chirp masses of the group of 15 low mass events that are reported to span $0.07<z<0.32$, with a mean of $z=0.17$, as listed in \cite{Abbott1+2} for periods 1\&2 and in \cite{Abbott2021} for period 3. The black hole masses comprising each binary are drawn in our model from a log-normal mass distribution in two ways, firstly by drawing each one randomly and secondly subject to the observed mass ratio of $1.45$ with a small dispersion in the component mass ratio of $10\%$, shown for comparison with the data in Figure~1 \& 2. We see this provides a very good reproduction of observed transition above the stellar black hole mass events as the transition from unlensed to lensed events and also matches the slowly rising behaviour in the high mass region above this transition, with the majority of high mass BBH events predicted to span $1<z<5$. This lens model also accounts for the bimodal appearance of the distribution of black hole masses shown in the histogram of Figure~3, with unlensed predicted events matching well the sharp low mass peak, whereas lensed events match well the broad distribution of high mass events peaking at $z\simeq 3$, corresponding to mean chirp mass of $40M_\odot$. The relative numbers of events between these two peaks allows us to determine that the intrinsic BBH event rate peaks at $z \simeq 3$, at a level that is about $\simeq 3$ orders of magnitude higher than the rate of unlensed detections at $z\simeq 0.2$,
but is well below the rate of formation of BHs implied by the rate of core SNe at $z\simeq 2$\cite{Diego_extreme} (see Appendix for rate evolution model details). 

   We also compare our lens model in the mass-distance plane shown in Figure~4, for the reported values of chirp mass and distance, i.e. with no account for lensing. This comparison demonstrates there is good consistency between the unlensed input events with $M < 20M_\odot$, in the stellar mass range, and close agreement with the higher chirp BBH events that we predict are lensed in the redshift range $1<z<5$, comprising 80\% of all reported BBH events to date.  
   
\begin{figure}[ht]
 \vspace{-1pt}
    \includegraphics[width=0.45\textwidth]{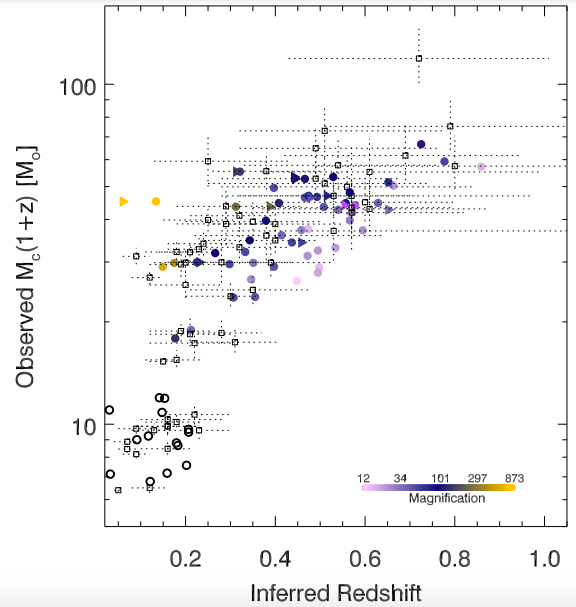}
 \vspace{-1pt}
\caption{{\bf Distance-Mass plane comparison.} This includes the most trusted reported BBH events marked with their measurement errors. The open and coloured circles are our unlensed and lensed event predictions respectively, where the colour indicates magnification. Colored triangles indicate a predicted second detection that is strong enough to be a repeat event. The predictions are made as if lensing is unrecognised, i.e. for comparison with the reported values. The simulated events correspond to 1 year of observation at O3 sensitivity. 
}
\end{figure}



 In addition to the dominant BBH population of uniform mass ratio identified above, there are a few events with much higher mass ratios visible in Figure~1 (marked by white lines) centered on $m_1/m_2\simeq 5$. Of these, the two with lowest chirp masses have been identified as low redshift NSBH events\cite{AbbottNSBH} with neutron star masses of 1.5 \& 1.9$M_\odot$ and with black hole masses of 5.7 \& 8.9$M_\odot$ reported, indicating a conventional stellar origin for the black holes in these NSBH binaries. The next two events with high mass ratios shown in Figure~1 have been classified as “mass-gap” (MG), as their observed secondary component masses, 2.7 \& 3.3$M_\odot$, are more massive than neutron stars and less massive than stellar black holes\cite{AbbottMG}, which in the context of lensing we have simply concluded are examples of lensed NSBH events at $z \simeq 1$, with conventional NS and BH masses. This lensing interpretation can also be applied to the reported ``mass-asymmetric" event \cite{AbbottMA} because of its large mass ratio (Figure~1), which in the context of lensing we have argued is simply a higher redshift NSBH event, at $z\simeq 3.5$ so the NS mass then appears to exceed $5M_\odot$ and thus misclassified as a black hole\cite{BDS:2020}. 
 
 Finally, we tentatively identify a possible population of 5 ``second generation events"  marked in purple in Figure~1, with mass ratios centered on $m_1/m_2 \simeq 2.0$. In the context of our lensing interpretation such events are anticipated, as the could comprise a typical remnant $\simeq 19M_\odot$ black hole (formed from  typical first generation pairs of 8 \& 11 $M_\odot$ seen in the ``unlensed" region of Figure~1), paired with a typical first generation black hole of $\simeq 10 M_\odot$, thereby resulting in a second generation event with $m1/m2 \simeq 2$, like the observed events marked in purple in Figure~1. Such second generation events should not be surprising in our model given the high early rate of BBH events that we have concluded in the context of lensing. Clearly, more data is required to determine whether there is a  such preferred mass ratio or whether these events are just the mass ratio tail of the general BBH distribution.

 \section{Discussion and Conclusions}

Over 90 pairs of merging black holes have now been reported spanning a wide range in chirp mass. These events now appear to be bimodal in terms of their observed chirp masses (see Figure~3) and also most of these BBH events have surprisingly similar component mass ratios across the full range of chirp mass (see Figure~1\&2), with a well defined mean value of $m_1/m_2=1.45 \pm 0.03$. Here we have shown that a simple alteration in interpretation, by invoking lensing, explains these features and accounts for the unusual events from the two mass gap regions as conventional stellar remnants (NS and BH) that are already observed in local group astronomy. Our conclusion that lensing dominates current GW events contrasts with other estimates where the input BBH mas distribution is based on the observed BBH events i.e. implicitly assuming that lensing is not significant, and because the observed mass distribution is so broad then lensed events are unrecognisable by mass because they seldom exceed the chirp masses of intrinsically wide mass function adopted \cite{Ng,Otto} and are thus outnumbered by the predicted rate of unlensed events. In our estimation the real BBH mass function is narrow, represented by only the lowest 20\% of chirp masses with $5-15M_\odot$, now clearly recognisable as a population in the BBH data, and hence we come to the opposite conclusion that lensing accounts for the majority of higher chirp mass events as high redshift, $1<z<5$, stellar mass black holes.

Our conclusion that BBH events are predominantly lensed is analogous to the brightest infrared galaxies detected in large sky surveys which are predominantly lensed by intervening galaxies into clear Einstein rings with a mean radius of $\simeq 0.85"$  \cite{Wardlow,Negrello,Bussmann},
demonstrating that lensing has magnified distant infrared galaxies (typically at $z\simeq 2$) to levels commonly exceeding the fluxes of the brightest unlensed galaxies, such that they dominate FIR detections in the first decade of flux\cite{Wardlow,Negrello,Bussmann}. 
It is conceivable that these same lensed galaxies are predominantly the source of lensed BBH events where massive binary stars are abundant, so that a positive cross correlation may be detected once positional triangulation is feasible for statistical samples of BBH events.

Magnifications ranging over $10-500$ are implied by our lens model for the majority of high chirp mass BBH detections, levels which we do know are achievable for small sources when projected close to lensing caustics, including the individually lensed stars recently discovered on the Einstein radius of lensing clusters with magnifications of several thousand reported in some cases\cite{Kelly,Venumadhav,Diego,Oguri}. These highly magnified stars limit the proportion of dark matter in primordial black holes insignificant $< 5$\%, as only modest micro-lensing is seen, consistent with the projected density of stars visible in the lensing clusters\cite{Diego,Oguri}, disfavouring the primordial black hole hypothesis for BBH events\cite{Bird,Carr}. Instead, this absence of ``LIGO-like" black holes may be regarded as support for our lensing interpretation of the high chirp mass events as no new class of intrinsically high mass black holes is required, just conventional stellar black holes that are lensed at high redshift.

In addition to the case presented here, we have also raised a time delay argument, finding that the time difference between pairs of BBH events with indistinguishable waveforms and sky location \cite{LVT2021}, is consistent with the  time delays between lensed QSO images that range from days to weeks (Diego, Broadhurst \& Smoot 2021\cite{Diego2021}. Our lens model predicts that repeat events should be detectable for a sizeable minority $\simeq 10\%-20\% $ of BBH detections, set by the Earth-rotating angular sensitivity of LIGO/Virgo\cite{Broadhurst2018} which limits detections to an overhead (and underfoot) band of the sky and typically misses counter images in the current configuration. It should also be appreciated that the relative fluxes of close, highly magnified images of lensed QSO's and SNe\cite{SNlens}, typically differ by a factor of 2 in brightness, implying substructure in lensing galaxies is common on sub-arcsecond scales\cite{Treu2019,Chan}, in which case the weaker counter image will typically fall below the GW detection threshold, as most GW events are near the detection limit, SNR $\simeq 8 $, with few exceeding a level of SNR $> 16$ required for detection of the weaker event to overcome a factor $\simeq 2$ flux anomaly, with credible examples proposed\cite{Dai_pair}.
 
Irrespective of lensing, the viability of the high masses inferred for BBH events above $\simeq 50M_\odot$ is challenged theoretically by the physical limit from pair instability\cite{Woosley} and also empirically there is a claimed deficit of high mass stars $>40M_\odot$ that are metal poor\cite{Schoot} in the SMC and similar shortages in the LMC and Milky Way, which if general would disfavour larger stellar progenitors required at low redshift to account for the reported high mass of BBH events. Instead, our conclusions appear qualitatively aligned with simulations of high mass star formation, made well before the GW detections, that predict a well defined process of fragmentation and accretion for generating close multiples of higher mass stars\cite{Krum}, qualitatively supporting the uniformity of the observed BBH component masses and our lensing based conclusion that most BBH events comprise pairs of conventional stellar mass black holes formed at early times.

An independent check on the viability of our lensing interpretation is anticipated using the stochastic GW background that is integrated over all redshift and binary mergers. This background is predicted to be higher in the case of lensing, particularly at low GW frequencies due to the relatively high rate of events at high redshift required by our lensing interpretation\cite{Suvodip}. Another complementary possibility is the direct detection of micro-lensing modulated GW waveforms \cite{Diego_ML,Otto} that would imply the presence of a macro-lensing galaxy hosting the micro-lenses.

Finally, we emphasise that the broad
black hole mass function derived from GW events is not independently supported by local black hole masses, where there is a complete absence of stellar black holes above $20M_\odot$. However, there is excellent agreement between the narrow local stellar mass black hole mass function and the sharp peak of
BBH events seen at the low mass end, as we have shown here. The main point to appreciate now is that lensing of these low mass events does solve this seemingly contradictory situation by allowing the detection of conventional stellar mass black holes at cosmological redshifts where their chirp masses are high like the majority of BBH detections. Lensing surely provides the simplest resolution of this apparent contradiction and implies that 80\% of the reported BBH events should be substantially revised downwards in mass and upwards in distance.

\section{ Appendix A: BBH event rate evolution}
This appendix describes the model for the rate of mergers as a function of redshift. This model is inferred from the observations and the assumption that events with inferred chirp masses above 20 are all gravitationally lensed.  
Our model consists of three basic ingredients;
i) A low rate of mergers at low redshift ($z<0.3$) which accounts for the events that are not being lensed. \\
ii) A large rate at redshift $z>2$, needed in order to compensate the low probability of lensing.\\
iii) A transition between the low redshift and high redshift\\

We consider the simple scenario where the the low redshift rates can be approximated by a constant, and the high redshift rate follows a functional form similar to the cosmic star formation rate, beyond the point where star formation peaks ($z \approx 2$). For the transition phase we adopt  an exponential form although other forms could be considered, such as a Gaussian model. The current data does not allow to constrain the transition well, but impose more clear limits on the low-z and high-z rates. 
For simplicity we adopt the following functional form for the rate model;
\begin{equation}
    R(z) = N \times S(z) \times M(z) 
\end{equation}
where $N\approx6\times 10^3$ Gpc$^{-3}$ yr$^{-1}$ is the normalization constant, $S(z)$ is proportional to the star formation rate given by the standard Madau \& Dickinson (2014) model (Eq. 15 in 2014, ARA\&A, 52, 415), 
\begin{equation}
    S(z) = \frac{(1+z)^{2.7}}{1+[(1+z)/2.9]^{5.6}} 
\end{equation}

and $M(z)$ is a modulation function which is constant and equal to 1 for $z>2$, i.e $M(z>2)=1$, and below $z=2$ is given by; 
\begin{equation}
    M(z<2)=\frac{2\times10^{-4}}{0.1+z^{1.9}} + e^{-\frac{|T(z)-T_{max}|}{T_e}}
    \label{Eq_Rate}
\end{equation}
with $T(z)$ the lookback time at redshift $z$, $T_{max}$ the corresponding lookback time at $z=2$, and $T_e$ the parameter controlling the speed at which the transition between the low-z and high-z rates take place. The first term in Equation \ref{Eq_Rate} compensates the function $S(z)$ at low-z resulting in $R(z)\approx Cte$ at low-z. Alternatively, one could simply replace the rate $R(z)$ at $z<0.3$ by a constant, in order to reproduce the observed number and distribution in the Mass-redshift plane of the not-lensed events. 

\begin{figure}[ht]
  \includegraphics[width=0.45\textwidth]{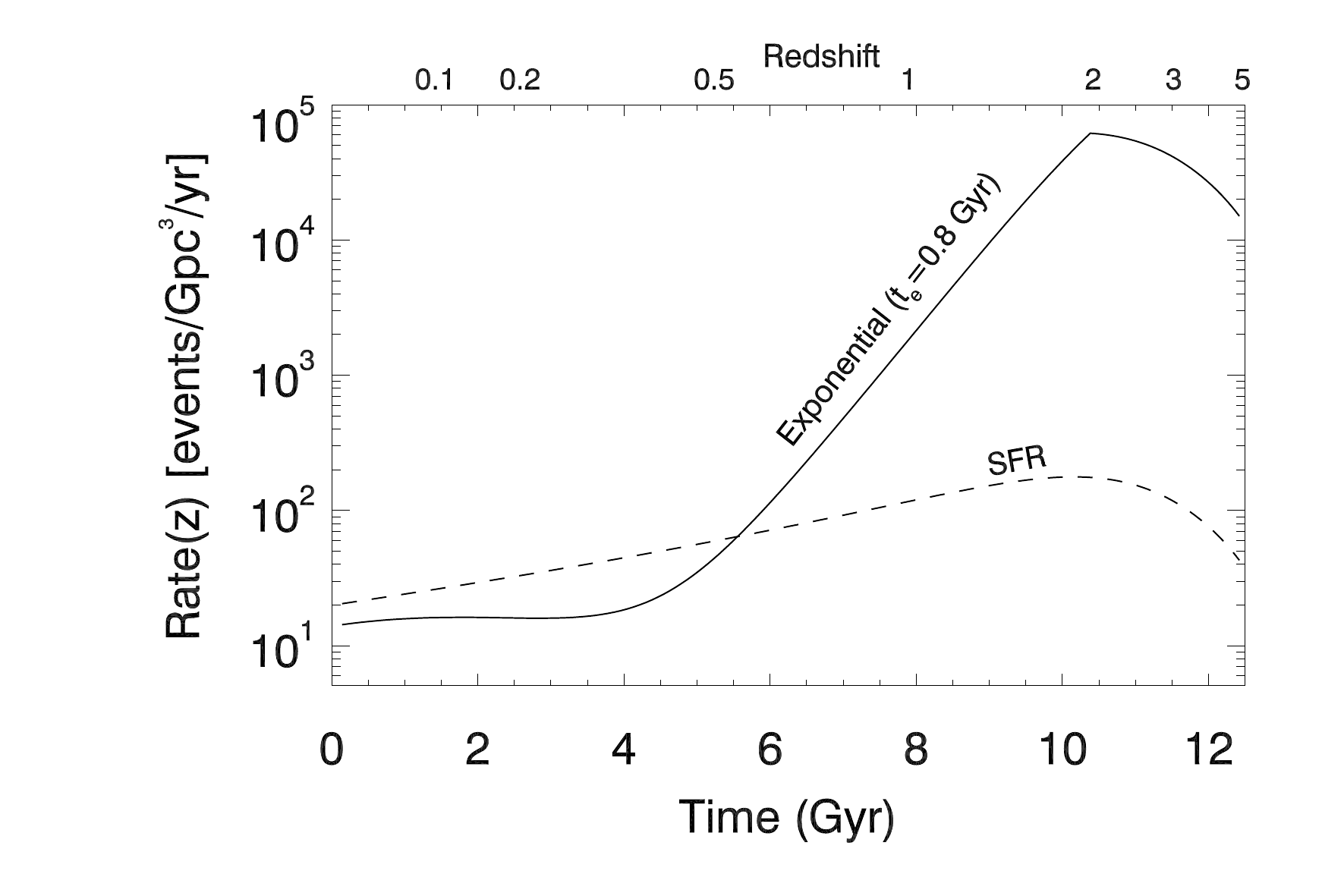}
\caption{{\bf Merger rate.} The solid line shows the implied merger rate from our lensing model, with $T_e=0.8$ Gyr. The dashed line shows a rate that traces the cosmic star formation rate, and is similar to the models assumed by LIGO under the assumption that no lensing is taking place.  
} \label{Fig_Rate}
\end{figure}

With the above assumptions the rate implied by our lensing hypothesis is shown in Figure \ref{Fig_Rate} as a solid line. In order to reproduce the observed distribution of mass and redshift for the low-z, low-Mass events, the local rate needs to be below the implied rate by LIGO (dashed curve in Figure \ref{Fig_Rate}), since under our lensing model hypothesis, some of the low-redshift events are in fact distant but amplified by lensing events. At high redshifts, our model implies a rate above $10^4$ events per year and Gpc$^3$. This rate is considerably higher than the one adopted by the LIGO collaboration under the no-lensing hypothesis. Such a high rate is needed in order to compensate for the low probability of lensing, but this rate is still well below the rate of SNe at these redshifts (or BH production). At a rate of $R(z=2)=5\times10^4$ Gpc$^{-3}$ yr$^{-1}$, ad adopting a galaxy abundance of 0.08 gal Mpc$^3$, consistent with UV the observed luminosity function at this redshift (see \cite{Diego2019}), we find that this rate implies 1 merger per year and per 1500 galaxies at $z=2$. Finally, our lensing interpretation favours a rapid decline in rate between $z \approx 2$ and $z \approx 0.3$, perhaps tied to the relatively rapid formation of massive star clusters.

\end{document}